\documentclass[a4paper,11pt]{article}
\usepackage{pos}

\usepackage{graphicx}
\usepackage{arydshln}

\newcommand{\ind}[1]{\rm\scriptscriptstyle #1}

\def\lsim{\mathrel{\rlap{\lower4pt\hbox{\hskip1pt$\sim$}}
    \raise1pt\hbox{$<$}}}                
\def\gsim{\mathrel{\rlap{\lower4pt\hbox{\hskip1pt$\sim$}}
    \raise1pt\hbox{$>$}}}                

\begin{document}


\title{
\vspace*{-0.75cm}
\begin{minipage}{\textwidth}
\begin{flushright}
\texttt{\footnotesize
PoS(LATTICE2021)490  \\
ADP-21-21/T1168      \\
DESY 21-213          \\
Liverpool LTH 1278   \\
}
\end{flushright}
\end{minipage}\\[15pt]
\vspace*{+0.75cm}
        Patterns of flavour symmetry breaking in hadron matrix elements 
        involving u, d and s quarks}

\ShortTitle{Patterns of flavour symmetry breaking \ldots}

\author[a]{J.~M. Bickerton}
\author[b]{A.~N. Cooke}
\author*[b,1]{R. Horsley}
\author[c]{Y. Nakamura}
\author[d]{H. Perlt}
\author[e]{D. Pleiter}
\author[f]{P.~E.~L. Rakow}
\author[g]{G. Schierholz}
\author[h]{H. St\"uben}
\author[a]{R.~D. Young}
\author[a]{J.~M. Zanotti}

\affiliation[a]{CSSM, Department of Physics,
               University of Adelaide, Adelaide SA 5005, Australia}
\affiliation[b]{School of Physics and Astronomy, University of Edinburgh,
                Edinburgh  EH9 3FD, UK}
\affiliation[c]{RIKEN Center for Computational Science,
                Kobe, Hyogo 650-0047, Japan}
\affiliation[d]{Institut f\"ur Theoretische Physik,
                Universit\"at Leipzig, 04109 Leipzig, Germany}
\affiliation[e]{PDC Center for High Performance Computing, 
                KTH Royal Institute of Technology, \\
                SE-100 44 Stockholm, Sweden}
\affiliation[f]{Theoretical Physics Division,
                Department of Mathematical Sciences,
                University of Liverpool, \\
                Liverpool L69 3BX, UK}
\affiliation[g]{Deutsches Elektronen-Synchrotron DESY,
                Notkestr. 85, 22607 Hamburg, Germany}
\affiliation[h]{Universit\"at Hamburg, Regionales Rechenzentrum,
                20146 Hamburg, Germany}

\note{For the QCDSF-UKQCD-CSSM Collaborations}

\emailAdd{rhorsley@ph.ed.ac.uk}

\abstract{Using an $SU(3)$-flavour symmetry breaking expansion between the
          strange and light quark masses, we determine how this constrains
          the extrapolation of baryon octet matrix elements and 
          form factors. In particular we can construct certain combinations, 
          which fan out from the symmetric point (when all the quark masses 
          are degenerate) to the point where the light and strange quarks 
          take their physical values. 
          As a further example we consider the vector amplitude at
          zero momentum transfer for flavour changing currents.}

\FullConference{%
 The 38th International Symposium on Lattice Field Theory, LATTICE2021
 26th-30th July, 2021
 Zoom/Gather@Massachusetts Institute of Technology
}

\maketitle


\section{Introduction and background}

\begin{figure}[!h]
\begin{minipage}{0.45\textwidth}

      \begin{center}
         \includegraphics[width=5.50cm]{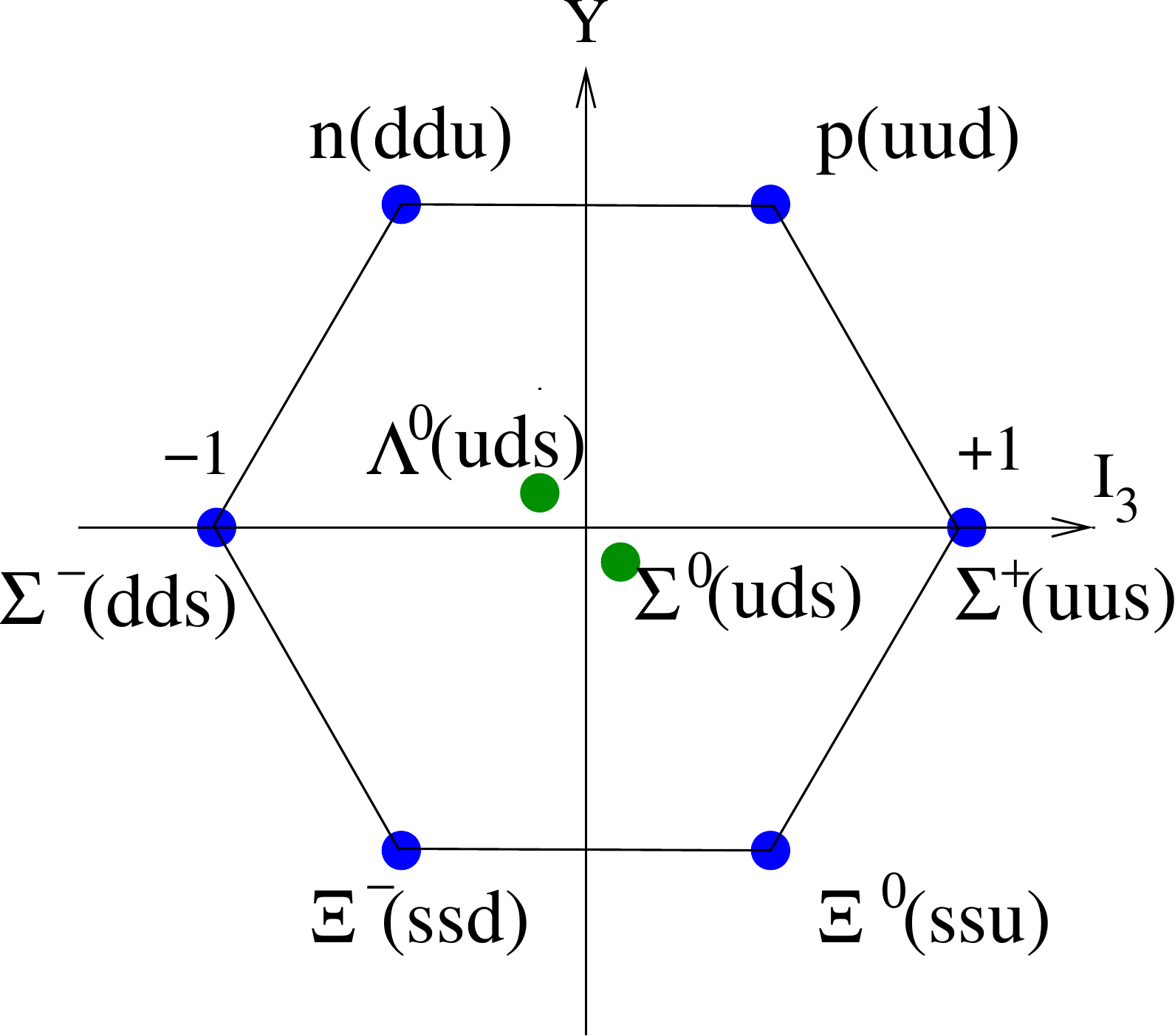}
      \end{center} 

\end{minipage}\hspace*{0.05\textwidth}
\begin{minipage}{0.50\textwidth}

      \begin{center}
         \includegraphics[width=5.00cm]{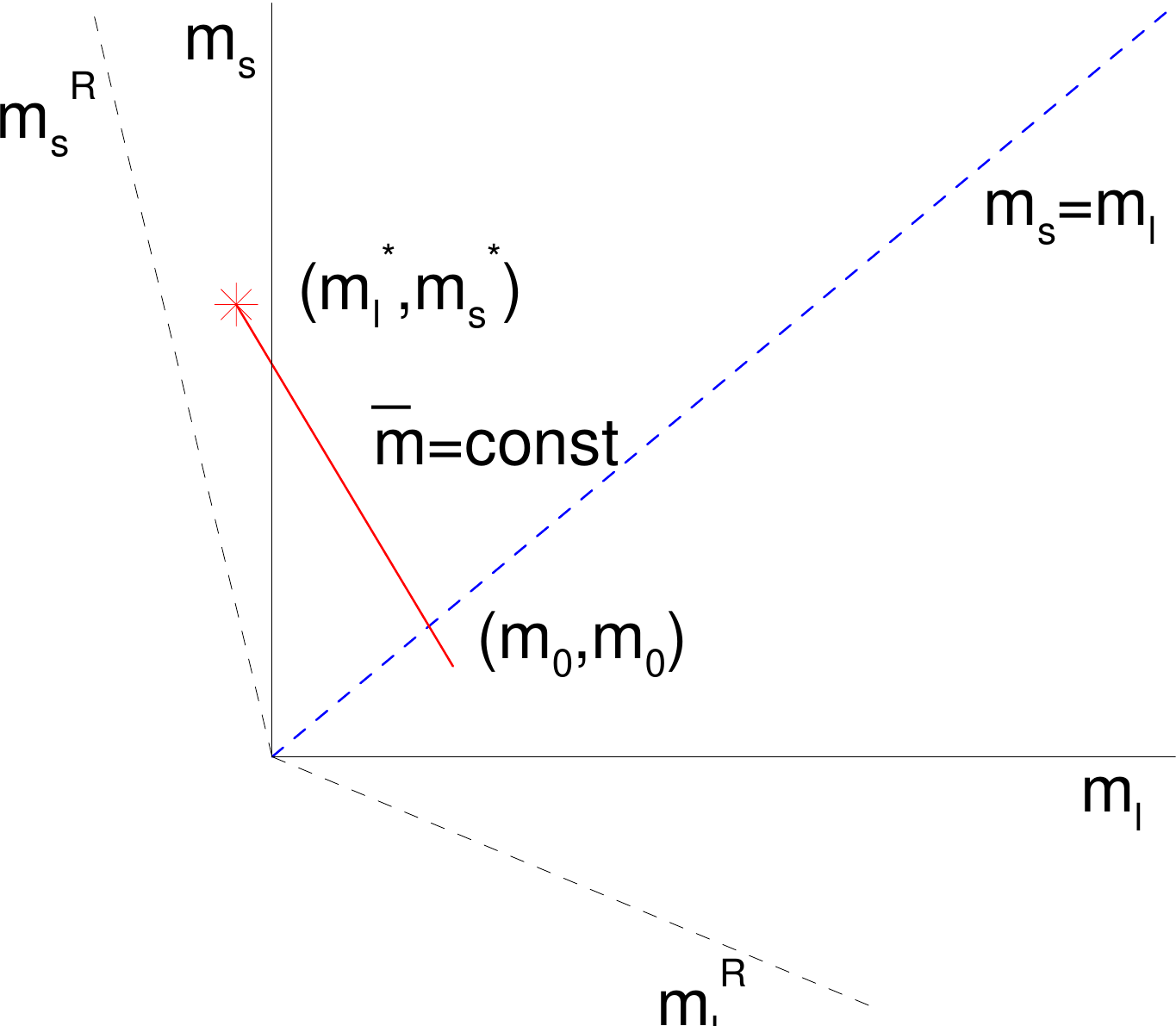}
      \end{center} 

\end{minipage}
\caption{Left panel: The baryon octet in the $I_3$--$Y$ plane.
         Right panel: The path from the $SU(3)$-flavour symmetry line
         to the physical quark mass (denoted by a star).
         The axes for the renormalised quarks are non-orthogonal
         which depicts the situation for non-chiral actions, 
         such as clover-Wilson fermions, 
         \protect\cite{Bietenholz:2011qq}.}
\label{path+octet}
\end{figure}

Understanding the pattern of flavour symmetry breaking and mixing, and
the origin of CP violation, remains one of the outstanding problems in
particle physics. Questions to be answered include
(i) What determines the observed pattern of quark and lepton
mass matrices and (ii) Are there other sources of flavour symmetry
breaking? In \cite{Bietenholz:2010jr,Bietenholz:2011qq}
we have outlined a programme to systematically investigate the pattern
of flavour symmetry breaking for QCD with three quark flavours. 
The programme has been successfully applied to meson and baryon masses 
involving up, down and strange quarks and has been extended to include
QED effects \cite{Horsley:2015eaa,Horsley:2015vla}. This article will 
extend the investigation to include baryon octet matrix elements as 
reported in \cite{Bickerton:2019nyz}. (In particular the baryon
octet is sketched in the LH panel of Fig.~\ref{path+octet}.)

The QCDSF strategy applied mainly to $2+1$ simulations 
(i.e.\ $\bar{m}_u = m_d \equiv m_l$) is to note that while there are 
many paths to approach the physical point, it is particularly favourable to
extrapolate from a point on the $SU(3)$-flavour symmetry line,
where all the quark masses are the same, to the physical point
$(\bar{m}_0, \bar{m}_0) \longrightarrow(m^*_l, m^*_s)$  keeping the singlet 
quark mass $\bar{m} \equiv (2m_l + m_s)/3$ constant, as illustrated
in the RH panel of Fig.~\ref{path+octet}.
This path is called the `unitary line' as we expand in 
both sea and valence quarks.

Defining $\delta m_q  = m_q - \bar{m}$ as the expansion paremter about
the $SU(3)$-flavour symmetric point $\delta m_q = 0$ then we find
Taylor expansions which at low orders in $\delta m_l$ have {\it constraints}
between the various octet masses. For example for the baryon octet we have
\begin{eqnarray}
   M_N         &=& M_0 + 3A_1\delta m_l + \ldots
                                                             \nonumber \\
   M_\Lambda    &=& M_0  + 3A_2\delta m_l + \ldots
                                                             \nonumber \\
   M_\Sigma     &=& M_0  - 3A_2\delta m_l + \ldots
                                                             \nonumber \\
   M_\Xi       &=& M_0 - 3(A_1-A_2)\delta m_l + \ldots \,.
\label{mass_expansion}
\end{eqnarray}
This expansion is known up to and including $\delta m_l^3$ terms. Thus
plotting the masses against $\delta m_l$, they fan out from the point
$\delta m_l = 0$. A further consequence is that singlet terms, such as
$X_N = (M_N + M_\Sigma + M_\Xi)/3 = M_0 + O(\delta m_l^2)$ have no 
$O(\delta m_l)$ terms. We shall find similar behaviour for the matrix
elements.


\section{Matrix element expansions}


We now develop similar expansions for matrix elements given by 
\begin{eqnarray}
   \langle B_i | J^{F_j} | B_k \rangle \equiv A_{\bar{B}_i F_j B_k} \,,
\end{eqnarray}
where our conventions are given in Table~\ref{ind8} for the possible
\begin{table}[!t]  
\begin{center} 
   \begin{tabular} {c|ccc}
      Index & Baryon ($B$) & Meson ($F$) & Current ($J^{\ind{F}}$) \\
      \hline
      \rule{0pt}{1.0\normalbaselineskip} 
      1     & $n$ & $K^0$ & $ \bar d \gamma s $  \\
      2     & $p$ & $K^+$ & $ \bar u \gamma s $  \\  
      3     & $\Sigma^-$ &  $\pi^-$ & $ \bar d \gamma u $  \\  
      4     & $\Sigma^0$ & $\pi^0$ & $ \frac{1}{\sqrt{2}}
                         \left(\bar u \gamma u - \bar d \gamma d \right) $ \\   
      5     & $\Lambda^0$&  $\eta$  & $ \frac{1}{\sqrt{6}}
                                        \left(\bar u \gamma u +
                         \bar d \gamma d -2  \bar s \gamma s \right) $ \\   
      6     & $\Sigma^+$ & $\pi^+$ & $\bar u \gamma d $ \\  
      7     & $\Xi^-$ & $K^-$ & $\bar s \gamma u $ \\  
      8     & $\Xi^0$ & $\bar K^0$ & $\bar s \gamma d $  \\
      \hline
      \rule{0pt}{1.0\normalbaselineskip}
      0     &         & $\eta^\prime$ & ${1 \over \sqrt{3}}
                        \left(\bar u \gamma u + \bar d \gamma d +
                               \bar s \gamma s \right)$  \\
\end{tabular}
\caption{Our numbering and conventions for the generalised currents.
         For example, $B_3 = \Sigma^-$, $F_3 = \pi^-$,
         $J^{F_3} \equiv J^{\pi^-} = \bar{d}\gamma u$. We use the convention 
         that current (i.e.\ operator) numbered by $i$ has the same effect 
         as absorbing a meson with the index $i$. $\gamma$ represents an 
         arbitrary Dirac matrix.}
\label{ind8}
\end{center} 
\end{table} 
octet states, $i = 1, \ldots, 8$ and the singlet state, 
labelled by $i=0$ (which is considered separately). As we are primarily 
concerned with the flavour structure of bilinear operators, 
we use the corresponding meson name for the flavour structure 
of the bilinear quark currents. So for example the $i = 5$ current
is given by the flavour matrix $F_{\eta} = \mbox{diag}(1,1,-2)/\sqrt{6}$. 
We shall use the convention that the current $i$ has the same
effect as absorbing a meson with the same index.
As an example, we note that absorbing a $\pi^+$ annihilates one 
$d$ quark and creates a $u$ quark. That is
$J^{\pi^+}|0\rangle \propto |\pi^+\rangle$. 
When $i \ne k$ we have transition matrix elements; when $i=k$
within the same multiplet, we have operator expectation values.
This has already been indicated in Table~\ref{ind8}.

In the case of $n_f = 2+1$ flavours considered here we only need
to give the amplitudes for one particle in each isospin multiplet,
and can then use isospin symmetry to calculate all other
amplitudes in (or between) the same multiplets. So, for example,
we can calculate the $\Sigma^-$ and $\Sigma^0$ matrix elements
if we are given all the $\Sigma^+$ matrix elements. Similarly,
given the $\Sigma^- \to n$ transition amplitude, we can find
all the other $\Sigma \to N$ transition amplitudes.

Within the set of amplitudes between baryons there are $7$ diagonal 
matrix elements:
$A_{\bar{N}\eta N}$, $A_{\bar{\Sigma}\eta \Sigma}$,
$A_{\bar{\Lambda}\eta \Lambda}$, $A_{\bar{\Xi}\eta \Xi}$ ($I=0$)
and $A_{\bar{N}\pi N}$, $A_{\bar{\Sigma}\pi \Sigma}$, $A_{\bar{\Xi}\pi \Xi}$ ($I=1$)
and $5$ transition amplitudes:
$A_{\bar{\Sigma}\pi \Lambda}$ ($I=0$), and $A_{\bar{N}K \Sigma}$, $A_{\bar{N}K \Lambda}$,
$A_{\bar{\Lambda}K \Xi}$, $A_{\bar{\Sigma}K \Xi}$ ($I=1/2$) giving $7+5 = 12$ in total. 
(There are a further $5$ inverse transition amplitudes, simply related
to the previous $5$ transition amplitudes.)
A Wigner-Eckart type theorem applies, the `reduced' matrix element 
(or amplitude) being multiplied by a Clebsch--Gordan coefficient. 
For example 
$\langle p |J^{\pi^+} | n \rangle 
       = \sqrt{2}\, A_{\bar{N}\pi N}
       =  \sqrt{2} \; \langle p | J^{\pi^0} | p \rangle$
giving
$\langle p | \bar u \gamma d | n \rangle 
      =  \langle p | ( \bar u \gamma u - \bar d \gamma d )
                            | p \rangle$.
Further details and tables are given in \cite{Bickerton:2019nyz}.

Matrix elements follow the schematic pattern for 2+1:
\begin{eqnarray} 
   \langle B_i | J^{F_j} | B_k \rangle 
      &=& \sum (\mbox{singlet mass polynomial}) \times 
                           (\mbox{singlet tensor})_{ijk} 
                                                        \nonumber  \\
      & & + \sum (\mbox{octet mass polynomial}) \times 
                           (\mbox{octet tensor})_{ijk} 
                                                        \nonumber  \\
      & & + \sum (\mbox{27-plet mass polynomial}) \times 
                           (\mbox{27-plet tensor})_{ijk} 
                                                        \nonumber  \\
      & & + \sum (\mbox{64-plet mass polynomial}) \times 
                           (\mbox{64-plet tensor})_{ijk} \,.
\end{eqnarray}
We already know the mass polynomials, \cite{Bietenholz:2011qq}, as
given in the LH panel of Table~\ref{su3_expansions}.
\begin{table}[!t] 

\begin{minipage}{0.45\textwidth}

   \begin{center} 
   \begin{tabular} {c|cccc}
   Polynomial & \multicolumn{4}{c}{$SU(3)$}                      \\
   \hline 
   $1$ & $1$ &  &  &                                             \\
   $\delta m_l$ &  & $8$ &  &                                    \\
   $\delta m_l^2$ & $1$ & $8$ & $27$ &                            \\
   $\delta m_l^3$ & $1$ & $8$ & $27$ & $64$                      \\
\end{tabular}
\end{center} 

\end{minipage}\hspace*{0.00\textwidth}
\begin{minipage}{0.50\textwidth}

\vspace*{-0.175in}

   \begin{center}
   \begin{tabular} {c|cc|cc}
   $SU(3)$ & \multicolumn{2}{c}{$T$, $1^{\rm st}$ class} & \multicolumn{2}{c}{$T$, $2^{\rm nd}$ class} \\
           & $d{\rm -like}$ & $f{\rm -like}$ & $d{\rm -like}$ & $f{\rm -like}$  \\
   \hline 
   $1$  & $d$ & $f$ &  &  \\
   $8$  & $r_1$, $r_2$, $r_3$ & $s_1$, $s_2$ & $t_1$, $t_2$ & $u_1$ \\
   $27$ & $q_1$, $q_2$ & $w_1$, $w_2$ &  $x_1$ & $y_1$ \\
   $64$ & $z$ &  &  &  \\
   \end{tabular}
   \end{center}

\end{minipage}

\caption{Left panel: All the quark mass polynomials up to $O(\delta m_l^3)$
         classified by symmetry properties. 
         Right panel: The relevant $17$ $T$-tensors, tabulated according to 
         first/second class and $d$- or $f$-type.}
\label{su3_expansions}
\end{table} 
So, for example, the $27$-plet part contains $O(\delta m_l^2)$ and
$O(\delta m_l^3)$ terms.
It remains to classify the $T_{ijk} = \langle B_i | J^{F_j} | B_k \rangle$
$3$-index tensors, $T_{ijk}$ and so need to look for a $SU(3)$ decomposition
of $8\otimes 8\otimes 8$. Here we give a very brief sketch.
We consider the tensor under $SU(3)$ rotations:
$T^\prime_{ijk} = U^\dagger_{ia} T_{abc} U_{bj} U_{ck}$
and in particular the change in $T$ under an infinitesimal transformation 
by a generator $\lambda^\alpha$. Using isospin constraints and known 
Casimir eigenvalues (solving $512$ equations with Mathematica) imply 
that there are $17$ independent tensors with most $T_{ijk}$ elements 
zero or $\sqrt{\rm integer}$. These can then be further classified, 
by first defining a reflection matrix, $R$, which inverts the outer ring
of the octet. The tensors can then be divided into first or second class
depending on the symmetry
\begin{eqnarray} 
   {\rm first \ class \ \ \ }  & &  T_{ijk} = + T_{kai} R_{aj} \,,
                                                      \nonumber  \\  
   {\rm second \ class} & &  T_{ijk} = - T_{kai} R_{aj} \,,
                                                      \nonumber
\end{eqnarray}
which interchanges $B_i$ and $B_k$ and transposes the flavour matrix $F^j$.
This corresponds to the Weinberg classification of currents into
first and second class, \cite{Weinberg:1958ut}, as discussed in
\cite{Bickerton:2019nyz}.

There is an additional classification by the symmetry when $R$ is applied
to all three indices
\begin{eqnarray} 
   d{\rm -like} & &  T_{ijk} = + R_{ia} T_{abc} R_{bj} R_{ck} \,,
                                                      \nonumber  \\ 
   f{\rm -like}  & &  T_{ijk} = -R_{ia} T_{abc} R_{bj} R_{ck} \,.
                                                      \nonumber
\end{eqnarray}
We find eventually that there are 
$17$ tensors: two singlets, eight octets, six $27$-plets, one $64$-plet
all contained in the $8\otimes8\otimes8$ decomposition. These are listed
in the RH panel of Table~\ref{su3_expansions}.

We are now in a position to give the polynomial expansions of the 
amplitudes to $O(\delta m_l^3)$. The same notation is used for the
tensor and its coefficient. For example considering 
$\langle p | J^\eta | p \rangle \equiv \langle B_2 | J^5 | B_2 \rangle = A_{\bar{N}\eta N}$ 
at say $O(\delta m_l)$ we have from LH
Table~\ref{su3_expansions} that it is octet. From RH
Table~\ref{su3_expansions} we see that for first class, it can contain
possible $r_1$, $r_2$, $r_3$ and $s_1$, $s_2$ tensor contributions.
Checking which tensors have a non-zero $252$ component gives
a coefficient $(r_1-s_2)$. 

Thus, as an example, we eventually find for a $1^{\rm st}$-class current
\begin{eqnarray} 
   \langle p | J^\eta | p \rangle
               &=& A_{\bar{N}\eta N}
                                                    \nonumber  \\
               &=& \underbrace{\sqrt{3} f - d}_{1} 
                   + (\underbrace{r_1  - s_2}_{8}) \delta m_l
                                                    \nonumber  \\
               & & + (\underbrace{\sqrt{3}f^{\ind{x}} - d^{\ind{x}}}_{1}
                   + \underbrace{r_1^{\ind{x}} - s_2^{\ind{x}}}_{8}
                   + \underbrace{9q_1 + 3q_2 + 3\sqrt{3}w_2}_{27})\delta m_l^2
                                                    \nonumber  \\
               & & + (\underbrace{\sqrt{3}f^{\ind{xx}} - d^{\ind{xx}}}_{1}
                   + \underbrace{r_1^{\ind{xx}} - s_2^{\ind{xx}}}_{8}
                   + \underbrace{
                        9q_1^{\ind{x}} + 3q_2^{\ind{x}} + 3\sqrt{3}w_2^{\ind{x}}}_{27}
                   + \underbrace{3\sqrt{3}z}_{64})\delta m_l^3 \,,
\end{eqnarray}
(by this we mean the relevant $1^{\rm st}$-class form factor) 
and as a further example for a $2^{\rm nd}$-class current
\begin{eqnarray}
   \langle n| J^{K^+} | \Sigma^- \rangle
               &=& A_{\bar{N}K\Sigma}
                                                   \nonumber   \\
               &=& ( \sqrt{2}t_2 +\sqrt{6}u_1 ) \delta m_l
                   + ( \sqrt{2}t_2^{\ind{x}} +\sqrt{6}u_1^{\ind{x}}
                   + \sqrt{5}x_1 + \sqrt{2}y_1) \delta m_l^2 \,.
\end{eqnarray}
Complete tables are given in \cite{Bickerton:2019nyz}.

It is natural to ask what do we gain for all this effort. The answer
is that the expansions are constrained, as can be easily ascertained by
counting the available parameters. For $1^{\rm st}$-class currents
there are $7 + 5 = 12$ possible amplitudes and we have
from the RH panel of Table~\ref{su3_expansions}
\begin{itemize}

   \item $O(1)$ has $2_1=2$ parameters

   \item $O(\delta m_l)$ has $5_8=5$ parameters

   \item $O(\delta m_l^2)$ has $2_1+5_8+4_{27}=11$ parameters

\end{itemize}
while at $O(\delta m_l^3)$ we have $2_1+5_8+4_{27}+1_{64}=12$ parameters 
and $12$ amplitudes, so there are no further constraints.
(The subscript denotes the representation given in the RH panel of 
Table~\ref{su3_expansions}.) Similarly for second-class currents 
-- there are now $5$ possible amplitudes, the expansion starts at
$O(\delta m_l)$ and we have
\begin{itemize}

  \item $O(\delta m_l)$ has $3_8=3$ parameters

\end{itemize}
while at $O(\delta m_l^2)$ we already have $3_8+2_{27}=5$ parameters, so again
there are no further constraints. So in all cases we only have constraints
at low orders in $\delta m_l$.

Alternatively we can construct linear combinations of amplitudes so
that we have only $d$- or $f$-terms in the expansion. For example we have
at $O(\delta m_l)$ a $1^{\rm st}$-class $d$ expansion can be constructed
\begin{eqnarray}
 D_1 \equiv - ( A_{\bar N \eta N} + A_{\bar \Xi \eta \Xi} ) 
 &=& 2 d - 2 r_1 \delta m_l                                 \nonumber \\
 D_2 \equiv A_{\bar \Sigma \eta \Sigma }
 &=& 2 d + (r_1 + 2 \sqrt{3} r_3) \delta m_l                 \nonumber \\
 D_3 \equiv{} - A_{\bar \Lambda \eta \Lambda }
 &=& 2 d - (r_1 + 2 r_2) \delta m_l                          \nonumber \\
 D_4 \equiv \frac{1}{\sqrt{3}} ( A_{\bar N \pi  N} - A_{\bar \Xi \pi  \Xi} )
 &=& 2 d -\frac{4}{\sqrt{3}} r_3 \delta m_l                  \nonumber \\
 D_5 \equiv A_{\bar \Sigma \pi \Lambda }
 &=& 2 d + ( r_2 - \sqrt{3} r_3) \delta m_l                  \nonumber \\
 D_6 \equiv \frac{1}{\sqrt{6}}   (A_{\bar N K \Sigma} + A_{\bar \Sigma K \Xi} )
 &=& 2 d + \frac{2}{\sqrt{3}} r_3 \delta m_l                 \nonumber \\
 D_7 \equiv - ( A_{\bar N K \Lambda} + A_{\bar \Lambda K \Xi} ) 
 &=& 2 d - 2 r_2 \delta m_l
\end{eqnarray}
just as for the masses as in eq.~(\ref{mass_expansion}) -- a `$d$-fan'.
We have $7$ lines, but only $3$ slope parameters, $r_1$, $r_2$, $r_3$, so 
the splittings are highly constrained. We can also construct quantities
that are constant at $O(\delta m_l)$, for example
$X_D \equiv ( D_1 + 2 D_2 + 3 D_4)/6 = 2 d + O(\delta m_l^2)$.
(These `averages' are not unique, here we just use the diagonal terms.)
Similarly for the $f$-fan: there are $5$ lines, but only $2$ slope 
parameters, $s_1$, $s_2$, so splittings are again highly constrained.
Examples of `fan' plots and `averages' for the vector current are
given in \cite{Bickerton:2019nyz}. For a recent example for the
tensor charge, see \cite{rose_lat21}.

As a further example consider the renormalised vector current ($J^r \sim V_4$) 
at $Q^2 = 0$. This simply counts the quarks (positive) and anti-quarks
(negative). So for the $7$ diagonal amplitudes, the results are constant
and known. This gives
\begin{eqnarray}
   A_{\bar{N}\eta N} = \sqrt{3} f \,, \,
   A_{\bar{\Sigma}\eta \Sigma} = 0 \,, \,\,
   A_{\bar{\Lambda}\eta\Lambda} = 0 \,, \,\,
   A_{\bar{\Xi}\eta\Xi} = -\sqrt{3}f \,, \,\,
   A_{\bar{N}\pi N} = f \,, \,\,
   A_{\bar{\Sigma}\pi\Sigma} = 2f \,, \,\,
   A_{\bar{\Xi}\pi\Xi} = f \,,
\label{A_diag_F_1_Q2eq0}
\end{eqnarray}
with $f = 1/\sqrt{2}$, $d = 0$. Note that because $d$ vanishes then 
$A_{\bar{\Sigma}\eta \Sigma}$ and $A_{\bar{\Lambda}\eta \Lambda}$ are identically zero.
The vanishing of the $O(\delta m_l)$ terms leads immediately
to the vanishing of all the coefficients (i.e.\ $r_1$, $r_2$, $r_3$
and $s_1$, $s_2$). This then also implies that the $5$ transition matrix
elements also have no $O(\delta m_l)$ terms. This is the content of
the Ademollo--Gatto theorem \cite{Ademollo:1964sr}. At the next order,
$O(\delta m_l^2)$, for the diagonal amplitudes we have $11$ parameters but
$7$ constraint equations, so we can solve for $4$ parameters, which we
take to be $q_1$, $q_2$, $w_1$, $w_2$. Substituting into the transition
amplitudes gives
\begin{eqnarray}
   A_{\bar{\Sigma}\pi \Lambda} 
                 &=& 0 + 40\left(q_1-{1\over 3}q_2\right)
                                                 \delta m_l^2
                                                             \nonumber   \\
   A_{\bar{N}K \Sigma} 
                 &=& -\sqrt{2}f
                     + \left(5\sqrt{6}(q_1+q_2)-5\sqrt{2}(w_1-w_2)\right)
                                                 \delta m_l^2
                                                             \nonumber   \\
   A_{\bar{N}K \Lambda} &=& -\sqrt{3}f
                      + \left(5(q_1 - {5\over 3}q_2) 
                         + 5\sqrt{3}(w_1+{1\over 3}w_2)\right)\delta m_l^2
                                                             \nonumber   \\
   A_{\bar{\Lambda}K \Sigma} &=& \sqrt{3}f
                      + \left(5(q_1 - {5\over 3}q_2) 
                         - 5\sqrt{3}(w_1+{1\over 3}w_2)\right)\delta m_l^2
                                                             \nonumber   \\
   A_{\bar{\Sigma}K\Xi} 
                 &=& \sqrt{2}f
                     + \left(5\sqrt{6}(q_1+q_2)-\sqrt{2}(w_1+w_2)\right)
                                                 \delta m_l^2 \,.
\label{A_transition_F_1_Q2eq0}
\end{eqnarray}
So we have one constraint between the $5$ amplitudes at $O(\delta m_l^2)$.

We finally note that to complete the job to determine all expansions,
we also need the singlet $\eta^\prime$ as given in Table~\ref{ind8}.
For example for 
$\bar{u}\gamma u = J^{\eta^\prime}/\sqrt{3} + J^{\pi^0}/\sqrt{2} + J^{\eta}/\sqrt{6}$
(and similarly for $\bar{d}\gamma d$, $\bar{s}\gamma s$) we need 
the singlet i.e.\ $A_{\bar{N}\eta^\prime N}$. But these expansions are 
$8\otimes1\otimes8$ and so have already been determined by the mass 
expansions \cite{Bietenholz:2011qq}. For example from 
eq.~(\ref{mass_expansion}) we have 
$A_{\bar{N}\eta^\prime N} = a_0 +3a_1\delta m_l + \ldots$ which allows for example
$\langle p|\bar{u}\gamma u|p\rangle$ to be determined. 
This construction is necessary for example for the electromagnetic current.
Again see \cite{Bickerton:2019nyz} for more details.


\section{Numerical results}


We consider $2+1$ Symanzik tree-level, $O(a)$ improved clover fermions,
\cite{Cundy:2009yy} at $\beta = 5.50$, where $a \sim 0.074\,\mbox{fm}$. 
At the flavour symmetric point $M_\pi \sim 465\,\mbox{MeV}$.

\begin{figure}[h]
   \begin{center}
      \includegraphics[width=14.00cm]
               {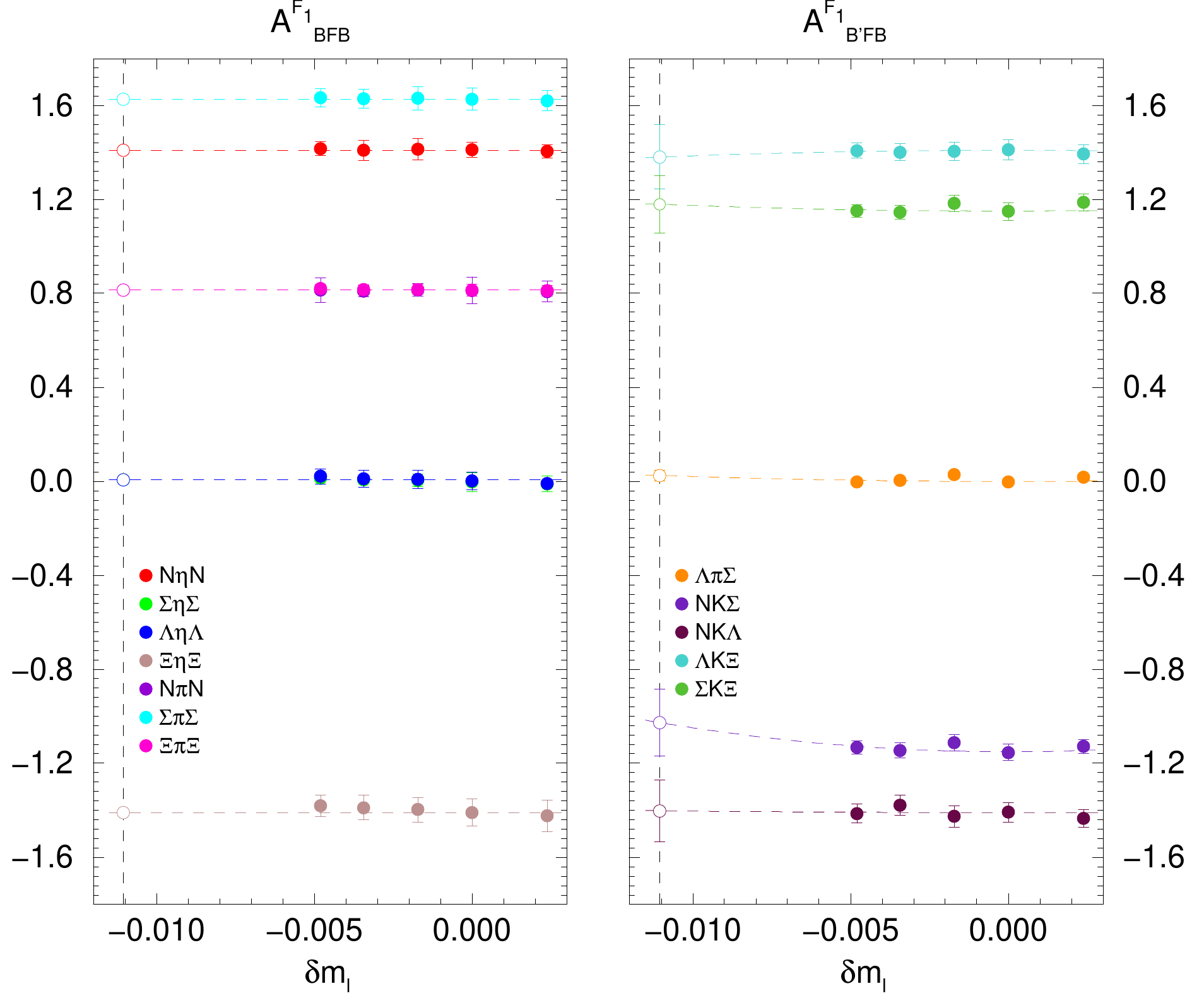}
   \end{center} 
\caption{Left panel: The $7$ diagonal $F_1$ amplitudes together with
         fits as described in the text. The filled points denote the
         numerical data, while the open points are at the physical
         pion mass. $\delta m_l = 0$ is the $SU(3)$-flavour symmetric
         point. Right panel: The $5$ transition $F_1$ amplitudes.}
\label{amplitude_all}
\end{figure}

For the vector current ($J \sim V_4$) we have determined the $12$
$F_1$ form factors (amplitudes) at $Q^2 = 0$ (using twisted boundary conditions
for the transition amplitudes to achieve this) on a $24^3\times 48$ lattice.
Our preliminary results (for $5$ quark masses) are given in 
Fig.~\ref{amplitude_all}.
Note that on the $\bar{m} = \mbox{const.}$ trajectory we have also
a data set with a light strange quark mass and a heavy
light quark mass. 

In the LH panel we show the diagonal amplitudes for $F_1$ and in the RH panel
the transition amplitudes also for $F_1$, together with a joint constrained
fit given by eqs.~(\ref{A_diag_F_1_Q2eq0}), (\ref{A_transition_F_1_Q2eq0}).
As expected first for the diagonal amplitudes the numerical data is 
very constant (i.e.\ independent of the quark mass), which the fit reproduces.
Also as the renormalised value of $f$ is $1/\sqrt{2}$, then presently
ignoring any $O(a)$-improvement, gives an estimate of the multiplicative
vector renormalisation constant, $Z_V$. As $A_{\bar{\Sigma}\eta \Sigma}$ and 
$A_{\bar{\Lambda}\eta \Lambda}$ are identically zero (i.e.\ have a zero fit 
function) we fit these separately with a constant, which gives a consistency
check of the data. 

Turning now to the RH panel of Fig.~\ref{amplitude_all} we show the 
transition amplitudes for $F_1$. The joint constrained fits are given 
by eq.~(\ref{A_transition_F_1_Q2eq0}). We find little evidence of 
discrepancies from the leading order, LO, constant values, except possibly for
$A_{\bar{N}K\Sigma}$. However as the fit coefficients presently have large 
uncertainties then it is really necessary to extend the computation to 
smaller quark masses, before any conclusion can be reached. 
It is interesting to note that \cite{Sasaki:2017jue} finds some evidence 
for discrepancies from the LO value for both $A_{\bar{N}K\Sigma}$ and 
$A_{\bar{\Sigma}K\Xi}$, however they both increase the absolute value 
of the amplitudes.


\section{Conclusions}


In this talk we have discussed baryon octet $SU(3)$-flavour symmetry breaking 
expansions for matrix elements (parallel to the previous mass expansions)
for $2+1$ quark flavours. This is complementary to chiral expansions
which start at a numerically out-of-reach zero quark mass, rather than 
here where we start at the $SU(3)$-flavour symmetry point. As for the mass 
case we again find constrained expansions. As an example, we have indicated 
that it might be possible to investigate discrepancies from the 
vector current LO values for the baryon octet at $Q^2 = 0$. 
Among various future extensions, one possibility is to consider the meson octet.


\section*{Acknowledgements}


The numerical configuration generation (using the BQCD lattice QCD 
program~\cite{Haar:2017ubh})) and data analysis (using the Chroma software 
library~\cite{Edwards:2004sx}) was carried out on the DiRAC Blue Gene Q 
and Extreme Scaling (EPCC, Edinburgh, UK) and Data Intensive (Cambridge, UK)
services, the GCS supercomputers JUQUEEN and JUWELS (NIC, J\"ulich, Germany)
and resources provided by HLRN (The North-German Supercomputer Alliance), 
the NCI National Facility in Canberra, Australia (supported by the 
Australian Commonwealth Government) and the Phoenix HPC service 
(University of Adelaide). RH is supported by STFC through grant ST/P000630/1.
HP is supported by DFG Grant No. PE 2792/2-1. PELR is supported in part 
by the STFC under contract ST/G00062X/1. GS is supported by 
DFG Grant No. SCHI 179/8-1. RDY and JMZ are supported by the 
Australian Research Council grant DP190100297.



\end{document}